\documentclass[aps,prl,showpacs,reprint,floatfix,preprintnumbers,nofootinbib]{revtex4-1}
\usepackage{amsmath,latexsym,amssymb,amsfonts,epsfig,xspace,amsthm}

\allowdisplaybreaks

\newcommand{\oiint}{\bigcirc \hspace{-1.35em} \int \hspace {-0.8em} \int}

\newcommand{\ket}[1]{\rvert #1 \rangle}
\newcommand{\norm}[1]{\left\lVert #1 \right\rVert}

\newcommand{\expec}[1]{\langle #1 \rangle}
\newcommand{\dm}{\text{d}\mu}

\newcommand{\pexp}[1]{\mathcal{P}\exp \oint_{#1}}

\newcommand{\pb}[1]{\underset{\Leftarrow}{#1}}

\newcommand{\lp}{\ell_\text{p}}
\newcommand{\spat}{\mathcal{S}}

\newcommand{\hilb}{\mathcal{H}}
\newcommand{\nj}{\,\rvert_j}
\newcommand{\njp}{\,\rvert_{j'}}
\newcommand{\njd}{\dashv_{j}}

\DeclareMathOperator{\conn}{\mathcal{A}}
\DeclareMathOperator{\tr}{tr}

\DeclareMathOperator{\ad}{ad}
\DeclareMathOperator{\Ad}{Ad}

\DeclareMathOperator{\one}{\boldsymbol{1}}
\DeclareMathOperator{\um}{\mathbb{I}}

\begin{document}

\date{1 Sep 2011}
\title{Chern-Simons expectation values and quantum horizons from LQG and the Duflo map}
\author{Hanno Sahlmann}
\email{sahlmann@apctp.org}
\affiliation{Asia Pacific Center for Theoretical Physics, Pohang (Korea)}
\affiliation{Physics Department, Pohang University for Science and Technology, Pohang (Korea)}
\author{Thomas Thiemann}
\email{thiemann@theorie3.physik.uni-erlangen.de}
\affiliation{Institute for Theoretical Physics III, University of Erlangen-N\"urnberg, Erlangen (Germany)}

\pacs{11.15.Yc, 04.60.Pp, 04.70.Dy, 02.10.Kn} 
\preprint{APCTP Pre2011-007}

\begin{abstract}
We report on a new approach to the calculation of Chern-Simons theory expectation values, using the mathematical underpinnings of loop quantum gravity, as well as the Duflo map, a quantization map for functions on Lie algebras. These new developments can be used in the quantum theory for certain types of black hole horizons, and they may offer new insights for loop quantum gravity, Chern-Simons theory and the theory of quantum groups. 
\end{abstract}
\maketitle
\section{Introduction}
\label{se_intro}
Loop quantum gravity (LQG) is based on a canonical formulation of gravity in terms of an SU(2) connection $A$ and a densitized triad field $E$ on a three dimensional spatial hypersurface $\spat$. Holonomies along paths $e$ and fluxes through surfaces $S$ in $\spat$, 
\begin{equation}
h_e[A]\equiv\pexp{e}A,\qquad E_{S,f}=\int_S *E^If_I 
\label{eq:}
\end{equation}
have well-defined operator analogues in the quantum theory \cite{Ashtekar:1996eg}. Here $*E$ is the 2-form
\begin{equation*}
(*E(p))_{ab}=\epsilon_{abc}\kappa^{IJ} E_I^c(p) T_J
\end{equation*}
dual to $E$, $\kappa^{IJ}$ is the inverse of the Cartan-Killing metric, and $T_{I}$ a basis of SU(2).  
It is apparent that the variables $A,E$ are treated on a slightly unequal footing in the quantum theory, and using different functionals of $E$ has been advocated in the literature, ex.\ \cite{Thiemann:2000bv,Fleischhack:2004jc}. Here we consider a proposal in the same direction. We will show that    
\begin{equation}
W_S=\mathcal{P}\exp 8\pi i c \oiint_{S}  \,*\widehat{E}^I\, \Ad_{h_s}(T_I)   \,\text{d}^2 s
\label{eq:w_def}
\end{equation}
can be well defined in the quantum theory. Here, $\widehat{E}$ is the LQG operator corresponding to $E$, $c$ is a real constant, the holonomies $h_s$ are along a path system from a chosen base-point on the boundary of $S$ to the point $s$, and the integral is \emph{surface ordered} as in the non-Abelian Stokes theorem \cite{Arefa:1980}. Motivation for the path ordering is a product law for the joining of surfaces, and simple behavior under gauge transformations. The $W_S$ are operator valued matrices with noncommuting entries. Intriguingly, despite their operator nature, they turn out to share many properties of SU(2) matrices, or rather those of some quantum deformation of SU(2), the precise nature of which is still to be determined. This emergence of quantum SU(2) in its kinematic setup is interesting for LQG in itself, but it also has interesting applications. On the one hand, we will sketch below that the simple condition 
\begin{equation}
h_{\partial S}\Psi=W_S\Psi  
\label{eq:qIH}
\end{equation}
for all surfaces $S$ lying in a surface $H$, seems to reduce a state $\Psi$ on $H$  to a solution of quantum 3d Euclidean gravity. Since  
\eqref{eq:qIH} is the quantum analogue of the \emph{horizon condition} for type I isolated horizons \cite{Engle:2009vc}, this shows a way to derive their quantum properties directly in LQG \cite{Sahlmann:2011xu}. On the other hand, we show how the operators $W_S$ can be used to calculate the Jones polynomial and its generalizations for certain simple links \cite{Sahlmann:2010bd,Sahlmann:2011uh}. 

A key ingredient in making \eqref{eq:w_def} well defined is an ordering procedure for the noncommuting components $\widehat{E}_I$. As we will explain below, it uses a fundamental structure from the theory of Lie algebras, the \emph{Duflo map} \cite{Duflo:1977}, and it is responsible for the occurrence of the quantum group structures. (The use of this map in LQG, albeit in a different context, was first suggested in \cite{Alekseev:2000hf}, and it has also been applied recently in \cite{Noui:2011im}.) This means that there is a connection between Duflo map and quantum groups that may open an interesting new perspective on the latter. 
\section{New flux operators}
\label{se_flux}
\paragraph{Definition.}
The basic ingredient in the definition of the operators $W_S$ are the flux operators \cite{Ashtekar:1996eg} of LQG. In the literature, these are always associated to a surface, but it turns out that the integrand exists independently as an operator valued distribution $\widehat{E}_I^a(s)$. It factors into two parts, 
$\widehat{E}_I^a(s)=\widehat{E}^a(s)\widehat{E}_I(s)$, which act on a holonomy
functional as follows: 
\begin{equation*}
\widehat{E}^a(s) h_e[A] = e^a(s) h_e[A], \quad e^a(s)=\int\text{d}t\, \dot{e}^a(t) \delta^{3}(s,e(t)), 
\end{equation*}
where $e(t)$ is the path along which $h_e$ is taken. Moreover
\begin{equation*}
\widehat{E}_I(s) h_e[A]= h_{e_2}[A] T_I h_{e_1}[A], 
\end{equation*}
where it was assumed that $s$ is the end point of $e_1$ and the beginning point of $e_2$ and $e=e_2\circ e_1$. If $s$ does not lie on $e$, the result of the action is the zero vector. Consideration of more general cases shows that $\widehat{E}_I(s)$ acts essentially like an invariant vector field $X_I$ of SU(2). Now, using these operator valued distributions we can define the operator $W_S$ as 
\begin{equation*}
\begin{split}
W_S
=\um_{2}
&+ 8\pi i c \int_S  
\Ad_{h_s}(*\widehat{E}(s))\\
&+\left(8\pi i c\right)^2
\int_{S^2} K_{s,s'}  
\Ad_{h_s}(*\widehat{E}(s))
\Ad_{h_{s'}}(*\widehat{E}(s'))\\
&+\ldots
\end{split}
\end{equation*}
Here, $K$ is an integration kernel that takes care of the surface ordering.
The holonomies $h_s$ connect the point $s$ on the surface with a base point on the boundary of $S$ via a system of paths in the surface, as in the non-Abelian Stokes theorem \cite{Arefa:1980}. There is some freedom in the choice of this path system, and the resulting operator will depend on it.  There are situations, however, in which the dependence on the path system drops out, see below. It is important to note that $*\widehat{E}$, when restricted to $S$, commutes with the holonomies along the path system as long as the surface has no self-intersections, due to the properties of $\widehat{E}^a$. We therefore restrict to that case in the following.    
Still, there are two problems with the above definition that have to be resolved. The first is that consecutive actions of $\widehat{E}^a(p)$ give rise to delta distributions that are concentrated precisely at the boundary of integration enforced by surface ordering, and a prescription for the evaluation of these has to be adopted. A straightforward regularization of the delta distributions
results in the prescription
\begin{equation*}
\int_{S^n} K_{s_1\ldots s_n} *e(s_1)\ldots*e(s_n)f(s_1,\ldots s_n)=\frac{1}{n!} f(s,\ldots s),
\end{equation*}
where the edge $e$ intersects $S$ in the point $s$.
The second problem with the above definition is that the operators $\widehat{E}_I(p)$ at a fixed point $p$ do not commute, whereas the components
$E_I$ of the classical field do. Therefore there is an ordering ambiguity inherent in the above definition. This ambiguity can be fixed using the {Duflo map}. This is a quantization map $\Upsilon$ from the free algebra of symbols $\{E_I\}_I$ with the Poisson bracket $\{E_I,E_J\}=f_{IJ}{}^KE_K,$
($f$ being the structure constants of a semisimple Lie-algebra $\mathfrak{g}$) into the universal enveloping algebra $U(\mathfrak{g})$, extending the map  $E_I\mapsto X_I$ on generators. The defining property of $\Upsilon$ is that it is an \emph{algebra isomorphism} between the invariant subspaces under the action of the corresponding Lie group $G$. 
$\Upsilon$ is an improved version of symmetric quantization $\chi$, 
\begin{equation}
\Upsilon = \chi\circ j^{\frac{1}{2}}(\partial), 
\label{eq:duflo}
\end{equation}
where $j^{\frac{1}{2}}(\partial)$ is a differential operator that can be obtained by inserting derivatives $\partial^I$ into the following function on $\mathfrak{g}$:
\begin{equation}
j^\frac{1}{2}(x)
=\det{}^\frac{1}{2}\left(\frac{\sinh\frac{1}{2}\ad x}{\frac{1}{2}\ad x}\right)
= 1 +\frac{1}{48} \norm{x}^2 +\ldots,
\label{eq:j12}
\end{equation}
with $\norm{x}^2=\tr(\ad_x^2)$ the square of the Cartan-Killing norm of $G$.
From now on, we will understand the powers of $*\widehat{E}$ to be ordered using $\Upsilon$.
For the calculations below, we will only need the action of $\Upsilon$ on the generators of the invariant subalgebra. For $G$=SU(2) we find 
\begin{equation}
\Upsilon (\norm{E}^2)= \Delta_{\text{SU(2)}} +\frac{1}{8}\one,  
\label{eq:duflo_basis}
\end{equation}
where $\Delta_{\text{SU(2)}}$ is the Laplacian. 

We finish the definition by considering the action of $W_S$ on the empty state $\ket{0}$. From the action of $\widehat{E}^a$ it is immediate that $W_S \ket{0}= \um_{2\times 2} \ket{0}$. It turns out, however that it is also useful to consider a different regularization, namely $W_S \ket{0}= \Upsilon\exp(8\pi c i E)\ket{0}= c_0 \um_{2\times 2}\ket{0}$ with $c_0$ a constant that is determined entirely by the shift in \eqref{eq:duflo_basis}. We will call this the \emph{alternative regularization} and compute $c_0$ below. 
 
\paragraph{Properties.}
In the following all the surfaces $S$ are 
oriented, simply connected, non-self-intersecting. They will also carry a path system that connects their base point to all the other points.
First of all, let $S_1+S_2$ be a surface that can be obtained as a disjoint union of two other surfaces, $S_1$ and $S_2$.  
Then, for specific choices of path systems on $S_1$, $S_2$, and $S_1+S_2$, such that they all share the base point, one can see that 
\begin{equation}
W_{S_1+S_2}=W_{S_1}W_{S_2}.
\label{eq:prod}
\end{equation}
If we use the alternative regularization for the action on spin networks without intersection, the above holds only on states that have at least one intersection in both, $S_1$, $S_2$. Due to the fact that $\Upsilon$ commutes with the action of $G$ one has
\begin{equation}
W^\dagger_S=W_{-S},
\end{equation}
where here and in the following $\dagger$ means taking both, transpose and operator adjoint for operator valued matrices. Finally, there is a unitary  action $U_g$ of gauge transformations on $\hilb$, and the operators $W_S$ transform covariantly, i.e., 
\begin{equation}
U_g\, W_S\, U_g^{-1}= g(b) W_S g(b)^{-1},
\end{equation}
where $b$ is the base point of $S$. 
 \begin{figure}
	\centerline{\epsfig{file=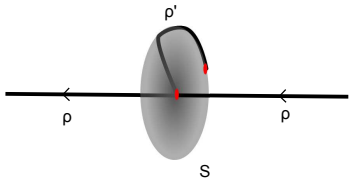,scale=.6}}
	\caption{New spin networks produced in the action of the operators $W_S$.}
	\label{fi:graph_change}
\end{figure}
We note that up to now, we have not made use of any specific properties of SU(2), and so everything remains valid for arbitrary semisimple gauge groups $G$.  
Let us now use $G=$SU(2), fix a surface $S$, and denote  by $\ket{\nj}$ any spin network state that has a single positively oriented transverse intersection with $S$ at an intersection point $p$. The spin on the edge intersecting $S$ is taken to be $j$. Taking the trace and considering only single intersections drastically simplifies the action of $W_S$: The holonomies along the path system drop out completely, and due to $\tr \circ \Upsilon=\Upsilon\circ \tr$ one obtains 
\begin{equation*}
\tr_j(W_S)\ket{\njp}=\Upsilon\tr_j \exp\left(8\pi i c E(p)\right)\ket{\njp}.
\end{equation*}
Then, using 
\begin{equation*}
\tr_j \exp a =\frac{\sin((2j+1)\theta/2)}{\sin(\theta/2)} \qquad \theta^2= -\frac{1}{2}\norm{a}^2
\end{equation*}
for an element $a$ of su(2), and \eqref{eq:duflo_basis} together with the isomorphism property of the Duflo map, one finds 
\begin{align}
\tr_j\left(W_{S}\right)\ket{\njp}
&=\frac{\sin\left[\pi c (2j+1)(2j'+1)\right]}{\sin\left[\pi c(2j'+1)\right]}\ket{\njp}\\
&=\tr_j(g_{j'})\ket{\njp}
\label{eq:ev}
\end{align}
with $g_{j'}=\exp(2\pi c (2j'+1)T_3)$ for $j'\neq 0$. For the alternative regularization, \eqref{eq:ev} stays valid for $j'=0$. Otherwise one has $\tr_j\left(W_{S}\right)\ket{0}=2\ket{0}$ and we redefine $g_0=1$. Either way, the last line shows that the eigenvalue can be written as the trace of an SU(2) group element. It follows that \emph{any} function of $\tr_j (W_{S})$, when acting on a state $\ket{j'}$ will have the properties expected of traces of an SU(2) element. This observation seems to even extend to the case of multiple intersections. One can for example show that for two adjacent surfaces $S_1,S_2$, each punctured once by a spin network $\ket{\nj \njp}$   
\begin{equation}
\begin{split}
\tr_{\frac{1}{2}}\left( W_{S_1}\right) &\tr_{\frac{1}{2}}\left( W_{S_2}\right)\ket{\nj \njp}\\
&=\tr_{\frac{1}{2}}\left(W_{S_1+S_2}\right)\tr_{\frac{1}{2}}\left( W_{S_1}W_{-S_2}\right)\ket{\nj \njp}
\end{split}
\label{eq:mandelstam}
\end{equation}
which is analogous to the property $(\tr g)(\tr g')=\tr(gg')\tr(g(g')^{-1})$ for group elements $g,g'$ of SU(2).
Finally, states $\ket{\nj}$ are not eigenstates of the operator valued matrices $W_S$. Rather,      
\begin{equation*}
W_S\ket{\nj}= c_1(j) \um_{2\times 2} \ket{j}+ c_2(j) \ket{\njd},  
\end{equation*}
where $c_1(j),c_2(j)$ are constants that are complicated to compute explicitly, and  
\begin{equation*}
\ket{\njd}= \Ad_{h_p}(T_I)\kappa^{IJ} \widehat{E}_J(p)\ket{\nj}
\end{equation*}
carries a new vertex at $p$ and a link between $p$ and the base point of the loop along the path system on $S$, see figure \ref{fi:graph_change}.  
\section{Applications}
\label{se_app}
\begin{figure}
	\centerline{\epsfig{file=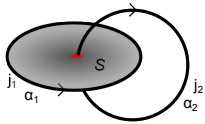,scale=.6}}
	\caption{The Hopf link spin network $H^+(j_1,j_2)$, and the surface $S$ with   $\partial S=\alpha_1$. }
	\label{fi:hplus}
\end{figure}

\paragraph{Chern-Simons theory.}
\label{se_cs}
Since the work of Witten \cite{Witten:1988hf} on Chern-Simons (CS) theory \cite{Deser:1982vy,Deser:1981wh}, it is well known that path integral expectation values of holonomy traces in CS theory are related to link invariants. For $G=$SU(2) and traces in the defining representation one obtains the \emph{Kauffman bracket}, with the conventional variable $A$ replaced by $q^{1/4}$, $q=\exp(2\pi i/k)$. (In fact, with CFT methods on finds $q=\exp(2\pi i/(k+2))$ \cite{Witten:1988hf}, but we do not see the shift of the level with our method.) We will now demonstrate that our new approach can reproduce some of these results. The key is that 
\begin{equation}
-\frac{8\pi  i}{k}\epsilon_{abc}\frac{\delta}{\delta A_c}e^{i S_{\text{CS}}[A]}= F_{ab}e^{i S_{\text{CS}}[A]}
\label{eq:func}
\end{equation}
for SU(2) CS theory with level $k$, and same, up to a numerical factor, for other gauge groups. 
 This has been exploited before \cite{Smolin:1989,Kauffman:1990,Gambini:1996mb}, but we will make use of an exponentiated version. The non-Abelian version of Stokes' theorem suggests that holonomy functionals can be replaced by the new flux operators under the path integral. To make this reasoning explicit, let $S$ be a smooth, oriented, simply connected surface, $\rho$ some representation of the structure group $G$, and $L$ be some functional of $G$-connections. Then formally  
\begin{equation}\begin{split}
\expec{L \tr_\rho(h_{\partial_S})}&=\int_{\conn} L[A] \tr_\rho(h_{\partial_S})[A] e^{i
S_{\text{CS}}[A]}\, \dm [A]\\
&=\int_{\conn}L[A] \tr_\rho(W_S)
e^{i S_{\text{CS}}[A]}\, \dm [A]\\
&=\int_{\conn} (\tr_\rho(W_{-S})L)[A]
e^{i S_{\text{CS}}[A]}\, \dm [A]\\
&=\expec{(\tr_\rho (W_{-S})L)}.
\end{split}
\label{eq:logic}
\end{equation}
While the manipulations under the path integral are formal, taking the first and the last line gives an equality in which all objects are defined, at least as long as $S$ has no self-intersections. Consider again $G$=SU(2) and choose $c=1/k$ in the definition of $W_S$. 
From the above, it follows, in particular, that the expectation value for unlinked traces of unknotted loops factorizes. The Kauffman bracket only factorizes if its value on the unknot is chosen appropriately. That means that to obtain the correct relationship with the Kauffman bracket, the expectation value of the trace for an unknotted loop must have a particular value. For SU(2), this is \emph{precisely} the case for the alternative regularization, thus we choose it in the following. 
To go further, let $H^+(j_1,j_2)=\tr_{j_1} (h_{\alpha_1})\tr_{j_2} (h_{\alpha_2})$ 
denote the right-handed Hopf link spin network, see figure \ref{fi:hplus}.
Applying \eqref{eq:logic} twice and using \eqref{eq:ev}, one finds 
\begin{equation}
\begin{split}
\expec{H^+(j_1,j_2)}_\text{CS}
&=\tr_{j_1}(g_{j_2})\tr_{j_2}(g_0)\\
&=\frac{\sin\left[\frac{\pi}{k}(2j_1+1)(2j_2+1)\right]}{\sin\left[\frac{\pi}{k}\right]},
\end{split}
\label{eq:verlinde}
\end{equation}
thus reproducing the known values for the Kauffman bracket and its generalization \cite{Martin:1989ub} for the framing induced by the choice of the surfaces (see figure \ref{fi:hplus}). 
Note that these numbers are important in related contexts: they are equivalently given by the trace of the square of the R-matrix of U$_q$(su(2)) on $j_1\otimes j_2$ or, up to normalization, by the Verlinde coefficients \cite{Verlinde:1988sn} in the SU(2)$_k$ WZW model of conformal field theory. 
Similar results can be obtained for other gauge groups, see for example \cite{Sahlmann:2011uh} for results on SU(3) traces in the defining representation.
\paragraph{Black hole horizons.}
\label{se_bh}
The quantization of an isolated horizon (IH) is a remarkable success of LQG \cite{Smolin:1995vq,Rovelli:1996dv,Ashtekar:1997yu,Kaul:2000kf,Domagala:2004jt,Meissner:2004ju,Corichi:2006wn,Engle:2010kt,Agullo:2010zz}. However, it is only an effective description, in the sense that it uses a number of elements that are not intrinsic to the formalism of LQG. For example, the location of the horizon is fixed to be the boundary of the space-time, and the fields on the boundary, although related to those in the bulk, are quantized separately, using a symplectic structure that is derived from the one on the bulk fields in the classical theory \cite{Ashtekar:2000eq,Engle:2010kt}. For spherically symmetric IH, the pullback to the IH, of the dynamical fields satisfy the horizon condition \cite{Engle:2010kt}
\begin{equation}
\pb{F}(A)=-\frac{\pi(1-\beta^2)}{a_H}\pb{\Sigma}(E).
\label{eq:horizon}
\end{equation}
In the quantum theory, we thus call a surface $H$ a type I quantum horizon, if \eqref{eq:qIH} is valid for all surfaces $S$ 
lying entirely within $H$. Here, the operators $W_S$ are evaluated with $c=-\pi\beta(1-\beta^2)\lp^2/2a_H$. 
Are there states that contain quantum horizons? Such states do not exist in the standard representation of LQG, but we argue that we can change the standard representation on $H$ in such a way that \eqref{eq:qIH} is satisfied. 
First, note that we can change the representation of the holonomies on $H$ without changing the representations of most other operators \cite{Sahlmann:2011xu}. Note now that a spin network $\ket{\psi}$ determines puncture data
\begin{equation*}
 \mathcal{P}=\{(p_1,j_1,m_1),(p_2,j_2,m_2),\ldots (p_N,j_N,m_N)\},
\end{equation*}
where $p_1 \ldots p_N$ are points on $H$ and $j_1 \ldots j_N$ and $m_1 \ldots m_N$ are labels of irreducible representations of SU(2), and magnetic quantum numbers in those representations. $\mathcal{P}$ defines a functional on functions of traces of loops that encircle at most one puncture:
\begin{equation}
\mu(\prod_i \tr_{k_i}(h_{\alpha_i})):= \prod_i \tr_{k_i}(g_{j_i}).  
\label{eq:su2def}
\end{equation}
This functional is positive \cite{Sahlmann:2011xu}, moreover we have seen above that it is consistent with all the relations among these traces. Does $\mu$ extend to all holonomy functionals on $H$? We think so. First of all, while we have not calculated traces of $W_S$ intersecting several loops, we have shown that the result must satisfy \eqref{eq:mandelstam} which, together with \eqref{eq:prod} generates many, if not all, relations among such traces.  
As for gauge noninvariant functionals, we may decompose them into a gauge invariant part and a functional on a tree graph, on which we use the standard measure. Changes in this decomposition should not matter due to the fact that 
$W_S=\um_{2\times 2}$ for $S$ not containing any punctures. The details are under investigation and will be reported elsewhere. We note that the resulting horizon theory seems to reproduce many of the results that have been obtained earlier \cite{Engle:2009vc,Engle:2011vf}, up to the fact that our results point to the theory being ISU(2) instead of SU(2) CS theory, i.e., Euclidean 3d gravity \cite{Sahlmann:2011xu}, due to the fact that on $H$ there remains one nontrivial holonomy \emph{and} one flux operator per nontrivial cycle of $H$.  
\section{Outlook}
\label{se_out}
In this Letter we have sketched the definition of new surface operators $W_S$ in the framework of LQG and their application to quantum CS theory and black hole horizons. There are a number of technical results that we would like to obtain, among them the extension of the action of $W_S$ to intersections at non-trivial vertices, and their definition for self-intersecting surfaces.   
These will make the operators even more relevant for knot theory. In particular we hope to apply our methods to the calculation of  Vassiliev invariants. As for applications to quantum IHs, a rigorous existence proof for the new representations is still outstanding, as well as a careful investigation of the physical consequences. We will furthermore consider the use of the operators $W_S$ in the SU(2) case as improved momentum operators in LQG, as they seem to possess many properties of (quantum) SU(2) group elements. Their use would thus reduce the asymmetry between configuration and momentum variables present in the standard parametrization. Finally  it appears that there is a deep relation between the Duflo map and quantum deformations of  Lie groups and algebras, which should be studied further. 

\begin{acknowledgments}
We thank Karim Noui and Lee Smolin for valuable conversations, and  
the organizers of the conference \emph{Loops 2011} in Madrid, where some of the ideas for this work originated. 
The research of HS was partially supported by the Spanish MICINN Project No.\ FIS2008-06078-C03-03. 
\end{acknowledgments}




\end{document}